\documentstyle[prd,aps,preprint,floats,epsfig]{revtex}

\tighten

\begin{document}
\draft
 
\pagestyle{empty}

\preprint{
\noindent
\today \\
\hfill
\begin{minipage}[t]{3in}
\begin{flushright}
LBNL--56862\\
UCB--PTH--05/01 \\
hep-ph/0501181 \\
January 2005
\end{flushright}
\end{minipage}
}

\title{Baryon helicity in B decay}

\author{Mahiko Suzuki}

\address{Department of Physics and Lawrence Berkeley National Laboratory\\
University of California, Berkeley, California 94720
}

\date{\today}
\maketitle

\begin{abstract}
  The unexpectedly large transverse polarization measured
in the decay $B\to\phi K^*$ poses the question whether it is
accounted for as a strong interaction effect or possibly points 
to a hidden nonstandard weak interaction. We extend here the
perturbative argument to the helicity structure of the 
two-body baryonic decay and discuss qualitatively on how 
the baryonic $B$ decay modes might help us in understanding 
the issue raised by $B\to\phi K^*$.  We find among others that
the helicity +1/2 amplitude dominates to the leading order in 
the $B$ ($\overline{b}q$) decay and that unlike the $B\to VV$ 
decay the dominant amplitude is sensitive to the right-handed 
$b\to s$ current, if any, in the penguin interaction.   
\end{abstract}
\pagestyle{plain}
\narrowtext

\setcounter{footnote}{0}
\section{Introduction}

In the two-body $B$ decay into light vector-mesons, the vector 
mesons should polarize longitudinally according to the 
simple $1/m_B$ power counting in the perturbative picture. 
The measured values of the longitudinal decay fraction $f_L$ 
are close to unity for the 
$\rho\rho$ modes\cite{Belle1,BaBar1,BaBar1'} in good agreement 
with this theoretical prediction. Quantitatively more reliable 
calculation can be made for $D^*\rho$ with the heavy-quark 
symmetry and the experimental value of $f_L\simeq 0.9$\cite{CLEO} 
agrees with theory\cite{Theory}. However, this simple prediction 
unexpectedly broke down for the decay $B\to\phi K^*$. The
value of $f_L$ turned out to be approximately 0.5 for 
$B\to\phi K^{*0}$;
\begin{equation}
       f_L = \left\{ \begin{array}{ll}
               0.43 \pm 0.09 \pm 0.04 \cite{Belle2},\\
               0.52 \pm 0.07 \pm 0.02  \cite{BaBar3}.
             \end{array} \right.
\end{equation}
The observed value for the charged mode $\phi K^{*+}$ is 
consistent with them; $f_L=0.46\pm 0.12 \pm 0.03$\cite{BaBar1}.
  As for the $\rho^+ K^{*0}$ modes, the situation is 
inconclusive at present since the numbers given by BaBar and 
Belle Collaborations are not perfectly consistent with each 
other; $0.79\pm 0.08\pm 0.04\pm 0.02$\cite{BaBar2} vs $0.50\pm 
0.19^{+0.05}_{-0.07}$\cite{Belle3}.
 
  Dominance of the longitudinal helicity is a direct 
consequence of the fact that the weak and strong forces are 
both mediated by gauge interactions, that is, 
chirality-conserving vector-axial-vector interactions. The 
longitudinal dominance should hold for all types of the decay 
interaction, either the tree or the penguin type, of the 
standard model.  In the 
limit that the light quark ($u,d,s$) masses are zero and the 
valence $q\overline{q}$ are collinear inside fast mesons, 
the longitudinal fraction $f_L$ would be unity for all of  
$\rho\rho$, $\rho K^*$ and $\phi K^*$. The difference 
between the $s$-quark in $\phi K^*$ and the $u/d$-quark 
in $\rho\rho$ should not be important if the strong interaction 
is strictly perturbative except at hadron formation. 

  There are two conceivable origins of the large transverse 
polarization in $B\to \phi K^*$. The first one is breakdown of
short-distance QCD dominance. That is, the strong interactions 
at long and/or intermediate distances may be somehow enhanced 
and cause helicity flip of quarks. For instance, if on-shell 
charm-anticharm meson intermediate states are important in 
the decay $B\to\phi K^*$, spins of slow charmed hadrons could 
flip with long-distance interactions and this effect would 
propagate into the light mesons in the final state\cite{Charming}. 
But our limited knowledge of dynamical parameters of the charm 
hadron sector makes a reliable estimate difficult. 
Another proposal has been made from the perturbative 
side: It was argued\cite{LeeSanda} that soft collinear quarks 
and gluons can enhance the annihilation decay process, which 
would be otherwise subleading in $1/m_B$. Although the spin 
flip cost another $1/m_B$, the soft-collinear loop corrections
in the annihilation decay might generate significant helicity 
flip in the case of $\phi K^*$\cite{Kagan}. While one can
parametrize such an effect, numerical estimate is subject to 
the uncertainty in the infrared and collinear cutoff. Yet 
another proposal is that the color-dipole decay operators may 
be nonperturbatively enhanced to generate a large transverse 
polarization\cite{Hou}. Many different proposals 
are being made to point to possible sources and mechanisms 
of the long-distance interactions responsible for the large
transverse polarization of $\phi K^*$.  However, it is not clear 
at present whether any of these proposals will really explain 
it as a strong interaction effect.

If the origin is not in strong interaction, a nonstandard 
decay interaction must be responsible. Is there a new decay 
interaction whose chirality structure is different from the 
standard gauge interaction ?  The case for such a new decay 
interaction is severely constrained. Fist of all, a new
interaction must be of the scalar-pseudoscalar or the tensor 
type.\footnote{
The $S$-$P$ interaction arising from Fierz rearrangement 
of the penguin operators does not solve the problem.}
It should couple preferentially with the $s$-quark if the 
problem exists only in $\phi K^*$, not in $\rho K^*$.
Furthermore the coupling should not have the quark-mass 
suppression $m_q/m_W$ unlike the standard Higgs coupling. 
While the possibility of the tensor weak coupling was pointed
out\cite{Kim}, it is yet to find a way to incorporate such an 
{\em ad hoc} interaction in the context of the electroweak 
gauge theory.  

  The fundamental issue is whether the breakdown of the helicity
rule is due to failure of the perturbative picture or to a
new weak interaction. If nonperturbative strong interactions are
responsible, how and where do they enter the decay processes? 
In addition to the pursuit from the theoretical side, more 
experimental information will help in reaching the root of the
problem. Study of the decay $B\to V(1^-)T(2^+)$ 
such as $B\to K^*f_2$ and $\phi K_2$ will be useful for this 
purpose. Indeed the first crude measurement of polarization has 
been made for the latter\cite{Andrei}. We call attention here
usefulness or relevance of the two-body baryon decays to 
the issue raised by the two-body meson modes. 
For instance, if large long-distance physics enters 
$B\to\phi K^*$ through the soft collinear corrections to the
annihilation process, the violation of the helicity rule would 
smaller in the corresponding baryonic decay modes since the 
annihilation decay is suppressed more severely for the baryonic 
decays than for the mesonic decays. As for the exotic decay 
interaction, one advantage of the baryonic decay over the mesonic 
decay is that the dominant helicity amplitude is sensitive 
to the right-handed current. The helicity rule in the baryon-pair
modes was discussed in an early paper by K\"{o}rner\cite{Ko}. 
He studied the baryonic decays with the tree interaction of $V-A$
using dynamical models for additional $q\overline{q}$ emission.
Now the penguin interaction is of our primary concern because of 
the $\phi K^*$ puzzle. We distinguish between the two processes 
here and present the results in a way relevant to the current 
issue of the $\phi K^*$ polarization. 

In the case of $B\to VV$, separation of the decay amplitudes 
into opposite-sign helicities $h=\pm 1$ requires
measurement of the $s/p$-wave interference between the
resonant $VV$ and the nonresonant $VPP$ background\cite{s-p} 
or a three-body decay correlation for some modes\cite{Pi} 
or else the angular correlation between decay products of 
different parents\cite{W}. 
In contrast, the helicity amplitudes of $h=\pm\frac{1}{2}$ 
in the baryonic decay can be easily separated with the angular 
analysis of a single hyperon in the final state if the decay 
violates parity. It is done as part of hyperon identification.   
Although the branching fractions of the baryon-pair modes are 
small according to early indications\cite{Belle4}, 
the simplicity in analysis will work to our advantage and 
allow us to accomplish the goal with much smaller samples of
data on the baryonic modes. 

  The paper is organized as follows:  After a brief review of 
the perturbative helicity selection rule for $B\to VV$ and its 
comparison with the data in Section II, we discuss the helicity 
rule for the baryon-antibaryon pair modes in Section III. In 
Section IV we discuss how to extract helicity information
from measurement and then select the baryonic modes that are 
useful for our purpose. We will not attempt detailed dynamical 
computation of the baryonic decay amplitudes since despite 
numerous theoretical efforts over years\cite{th1,th2} the
results are numerically less reliable for the baryonic modes   
than for $B\to VV$.  Instead we quote only semiquantitative 
estimates 
which are based primarily on simple perturbative dynamics
and symmetry, not on the specific form factors or the value of 
$\overline{\alpha}_s$. Such crude estimates are in good 
agreement with experiment for $B\to VV$ other than $\phi K^*$ 
and $\rho K^*$. In Section V, we summarize our results and 
discuss prospects in theory and experiment.
 
\section{Perturbative counting rule for meson pairs}
     
  The perturbative helicity rule in B decay is based on two 
facts of the standard model. First, the weak and strong 
interactions are both gauge interactions so that, whenever 
a light quark pair is produced, its chirality is given by 
$\overline{q_R}q_R$ or $\overline{q_L}q_L$, not 
$\overline{q_R}q_L$ nor $\overline{q_L}q_R$.\footnote{ 
In the perturbative power counting, the light-quark-pair 
production through the color-magnetic decay operator 
$\propto\overline{b_R}\sigma_{\mu\nu}G^{\mu\nu}q_L$ picks
up $O(1/m_B)$ through the light quark pair production
$\overline{q_L}\sigma^{\kappa\lambda}G_{\kappa\lambda}q_R
+(R\leftrightarrow L)$.}
The energetic quarks may be produced either directly by the 
decay interaction or through the hard gluon interaction.
The quark chirality does not change by emission nor absorption 
of hard gluons. Secondly, final hadron states are formed in the 
leading order by superposition of valence quarks with the 
light-cone wavefunctions. Therefore, helicity of a fast hadron 
is determined by helicities of its energetic constituents, 
$\overline{q}q$ for mesons and $qqq$ for baryons. The terms 
neglected in this approximation are of higher orders in 
$1/m_B$ or of higher-twist contributions in terms of the 
wavefunctions and effective operators. Breakdown of the 
helicity prediction therefore means that some long and/or 
intermediate distance strong interaction is enhanced to
overcome the power suppression of $1/m_B$. 

   Under these conditions the chiral content of the energetic 
quarks produced in the final state of $B\to VV$ is: 
\begin{equation}
    \sim (\overline{q_L}q_s)(\overline{q_L}q_L) \;\;{\rm or}\;\;
           (\overline{q_L}q_s)(\overline{q_R}q_R), \label{int}
\end{equation}
where $q$ stands for the quark state of $u,d,s,c$, the subscript 
of $q_s$ stands for the ``spectator''. It is understood that the
colors are saturated appropriately. By parity invariance, $q_s$ 
has equal probabilities of spin up and down. The chiral content 
of Eq. (\ref{int}) would not change in the limit of $m_q\to 0$ and 
$m_V\to 0$ even after any number of hard QCD interactons may take 
place. Eq. (\ref{int}) gives the chiral content of the valence
quarks/antiquarks of $VV$ not only for the spectator decay 
processes but also for the annihilation and exchange decay 
processes.  

To derive the helicity rule, consider the decay,
\begin{equation}
 B(\overline{b}q_s)\to V_1(\overline{q}q)+V_2(\overline{q}q_s).
\end{equation}
If $q_L$ and $\overline{q_L}$ fly in parallel to form one vector 
meson, this meson $V_1$ ($\overline{q_L}q_L$) is in the helicity 
state of $h=0$. In the other meson $V_2$, the $\overline{q_L}q_s$ 
pair alone can make $h=0$ or +1 since the spin of $q_s$ can point 
to either direction. But requirement of the overall $J_z=0$ forces 
the $V_2$ helicity to $h=0$ in this case. (See the first figure
in Fig.1.)  The same argument holds in the case of 
$V_1(\overline{q_R}q_R)$.

\noindent
\begin{figure}[h]
\epsfig{file=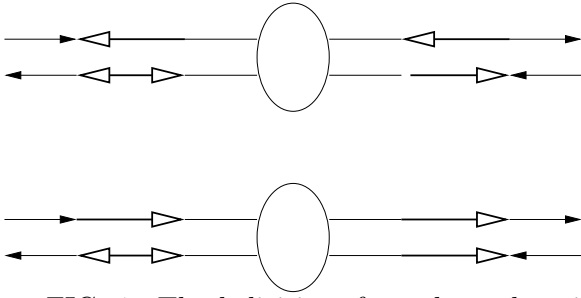,width=0.47\textwidth}
\caption{The helicities of quarks and antiquarks in
$B\to VV$ for $\to\overline{q_L}q_L+\overline{q_L}q_s$
(the upper figure) and for $\to\overline{q_L}q_R +
\overline{q_R}q_s$ (the lower figure). The solid arrows indicate
the quark-number directions, and the large open arrows stand
for the dominant helicities. The two-end open arrow is for $q_s$.
\label{fig:1}}
\end{figure}

Alternatively, if $q_R$ and $\overline{q_L}$ try to form $V_1$, 
helicity of $V_1$ ($\overline{q_L}q_R$) is in $h=+1$, But 
helicity of $V_2$ ($\overline{q_R}q_s$) can be only $h=0$ 
or $-1$, not $+1$ (the second figure in Fig.1). This conflicts 
with $J_z =0$. Therefore one concludes that 
the only allowed helicity state is the longitudinal ($h=0$) 
state for $V_1V_2$. The kinematical corrections to this rule 
arise in $O(1/m_B)$ from the transverse motion of $q\overline{q}$ 
inside a meson and the nonvanishing quark masses. Computation of 
these higher-twist terms can be carried out for $B\to VV$ by 
the QCD factorization method. In the case that one of the final 
mesons is a charmed meson, the form factor can be computed 
reliably with the heavy quark symmetry. In the case of light 
meson pairs, final results involve larger uncertainties 
due to the light-cone wavefunctions and the value of
$\overline{\alpha}_s$. Without going through this computation, 
however, a reasonable order-of-magnitude estimate can be made 
as we shall do below.

  The $h=+1$ amplitude is realized by the small {\em wrong} 
helicity component of $q_L$ in $V_1$ when 
$\overline{q_L}q_L$ form $V_1$ as $\overline{q_L}(x{\bf p})
(\gamma_1-i\gamma_2)q_L((1-x){\bf p})$ and the right-chiral 
component of $q_s$ combines $\overline{q_L}$ to form $V_2$ as 
$\overline{q_L}\sigma^{\mu\nu}q_s$. For the $h=-1$ amplitude, 
the wrong helicity component is needed for $\overline{q_L}$ 
in $V_1$ and also for $\overline{q_L}$ in $V_2$. According to 
the chiral projection 
of the plane Dirac wave, the wrong helicity component is 
suppressed by $m_q/(E_{{\bf k}}+|{\bf k}|)$. The transverse 
motion of $\overline{q}q$ inside a meson also acts as an 
effective quark mass under the longitudinal Lorentz 
transformation. Consequently the effect of the internal motion 
on the helicity can be incorporated by replacing the (current) 
quark mass $m_q$ with the transverse quark mass
$m_T=\sqrt{m_q^2 +{\bf k}_T^2}$.  We are thus led to the 
well-known hierarchy of the helicity amplitudes 
$H_{hh}$ for $B\to V_1({\bf p}h)V_2(-{\bf p}h)$
\cite{Theory,Counting};\footnote{
Interchange as $H_{++}\leftrightarrow H_{--}$ for the 
$\overline{B}(b\overline{q})$ decay.}
\begin{equation}
  \biggl|\frac{H_{++}}{H_{00}}\biggr|\simeq 
  \biggl|\frac{H_{--}}{H_{++}}\biggr|\simeq 
    \biggl\langle\frac{m_T}{E_{{\bf k}}+|{\bf k}|}\biggr\rangle,
     \;\;\;(m_T \ll E_{{\bf k}}),    \label{hierarchy} 
\end{equation}
where the bracket $\langle$ $\rangle$ denotes the average 
over the quark momentum with the light-cone wavefunction.
It is a reasonable approximation to set $m_T\simeq\frac{1}{2}m_V$ 
for the light mesons or a little more accurately $m_T\simeq 
\sqrt{\frac{2}{3}}\times\frac{1}{2}m_V$ by taking account of 
$\langle {\bf k}_T^2\rangle = \frac{2}{3}\langle{\bf k}^2\rangle$.
With $\langle E_{{\bf k}}\rangle \approx \frac{1}{2}E_V\approx
\frac{1}{4}m_B$, Eq. (\ref{hierarchy}) is a counting rule 
in $1/m_B$ based on kinematics. It applies to decay
amplitudes of a given decay operator. A total amplitude may be
sum of terms from different operators in general.
Instead of going through detailed dynamical calculation, we 
proceed here with a semiquantitative estimate. Let us
substitute $|{\bf k}|$ with its peak value of distribution
$\frac{1}{2}|{\bf p}|$. Then we obtain with Eq. (\ref{hierarchy}) 
the magnitude of the longitudinal fraction 
$f_L\equiv |H_{00}|^2/\sum_h|H_{hh}|^2$ as
\begin{equation}
    f_L \simeq  \left\{ \begin{array}{ll} 
           0.98\sim 0.99 & \;\;(\mbox{$\rho\rho$, $\rho K^*$}), \\
           0.96 & \;\;(\mbox{$\;\phi K^*$}).          
   \end{array} \right. .          \label{fL}
\end{equation} 
If the contribution of the end points of the wavefunctions 
is enhanced, these numbers can deviate more from unity. 
They are in line with measurement for the tree-dominated 
$\rho\rho$, off by two standard deviations or more on the 
larger side for the penguin-dominated $\rho K^*$, and clearly 
far too large for $\phi K^*$ which is expected to be almost 
purely a penguin decay.\footnote{ 
The same estimate leads to $f_L\simeq 0.92\sim 0.93$ for 
$D^*\rho$.  This number includes the $c$-quark mass effect 
(with the constituent mass $m_c\simeq 1.7$ GeV) in the 
left-chiral $\overline{c}$-quark state in $\overline{D}^*$.
The value of $f_L$ does not deviate much from unity since 
it is the wrong helicity of  $u_L/d_L$ in $\rho$ not of
$\overline{c}_L$ in $\overline{D}^*$ that is needed to realize 
$h=+1$ for this tree-decay process. It is in reasonable 
agreement with experiment, $0.890\pm 0.018\pm 0.016 
(\overline{D}^{*0}\rho^+)$ and $0.885\pm 0.016\pm 0.012 
(D^{*-}\rho^+)$\cite{CLEO2}.}
If a longitudinal helicity amplitude consists of more than 
one term and large cancellation occurs between different 
terms, the ratio of the transverse-to-longitudinal 
amplitude could be enhanced. We would need suppression of
factor five for $H_{00}$ to explain $f_L\simeq 0.5$ for 
$\phi K^*$ by such cancellation, which would translate to 
suppression of the 
$\phi K^*$ branching fraction by a factor of 25 relative 
to the case without cancellation. The observed values of 
the penguin decay branching fractions are not by  
order-of-magnitude off the conventional theoretical estimate. 
Therefore, it is not easy to attribute the observed large 
transverse polarization particularly in $\phi K^*$ to 
a strong suppression of the dominant longitudinal amplitude
by cancellation:
  
It is tempting to attribute the large transverse polarization 
of $\phi K^*$ to a new interaction of an unconventional chiral 
structure hidden in the penguin loop. However,
as long as the strong interaction dynamics
is of short distances, the right-hand weak current would not
solve the problem. Because the only difference of the
right-handed weak interaction from the left-handed weak 
interaction is to interchange $H_{++}\leftrightarrow H_{--}$ 
in Eq.(\ref{hierarchy}). To violate the 
helicity selection rule of Eq. (\ref{hierarchy}), such 
a new interaction must emit a quark pair through the $S$-$T$-$P$ 
interaction, $\overline{q_R}q_L\pm\overline{q_L}q_R$, instead of 
$V$-$A$ interaction, $\overline{q_R}q_R\pm\overline{q_L}{q_L}$.
The effective $S$-$T$-$P$ interaction from the 
Fierz-rearrangement of the left-right current interaction 
does not help since the helicity 
argument at the beginning of this Section can be made equally 
well for the interaction prior to the Fierz rearrangement. 
Because the coefficients of $S$-$T$-$P$ are fixed in such a
case so that only $H_{00}$ survives after the $S$-$T$-$P$ 
contributions are summed over.
Furthermore, if one has to explain that the transverse 
polarization is more pronounced in $\phi K^*$ than in $\rho K^*$, 
the new interaction should affect more strongly on $s$-quark 
than on $u/d$-quark. The Higgs interaction indeed follows such 
a coupling pattern, but the magnitude of the standard Higgs 
coupling is far too small to be relevant. 

\section{Baryon-antibaryon modes}   

    When a baryon-antibaryon pair ${\cal B}\overline{{\cal B}}$
is produced in $B$ decay with the four-quark decay interaction,
an additional pair of
$q_R\overline{q_R}$ or $q_L\overline{q_L}$ must be produced through
strong interaction. Since our interest is to test the
perturbative picture of the helicity structure for the
${\cal B}\overline{{\cal B}}$ modes, we consider the case where
the gluon producing the $q\overline{q}$ pair is hard and highly
virtual as much as $\sqrt{q^2}=\frac{1}{3}m_B$ on average. This costs
$\overline{\alpha}_s$ suppression, which is unavoidable and common
to all perturbative ${\cal B}\overline{{\cal B}}$ production.
K\"{o}rner noticed\cite{Ko} that in the tree interaction of $V-A$,
helicity mismatch occurs when the hard quark-antiquarks
enter ${\cal B}\overline{{\cal B}}$ as valences directly from
the primary decay interaction. He derived selection rules 
assuming no subsequent hard bend of the primary momenta. The
helicity mismatch does not occur for the penguin interaction 
since it contains $(V-A)(V+A)$. Once a hard gluon emission is
considered, however, the tree decay process can be saved from 
the helicity conflict under a certain circumstance: Imagine
that the hard gluon emission reverses the primary quark momentum
and takes away one unit of helicity. In this case the primary quark 
(or antiquark) momenta need not be parallel at the time of 
emission from the weak interaction. Then the selection rule 
of \cite{Ko} is evaded. A closer inspection of the matrix element 
proves that the tree-decay amplitudes are indeed allowed even 
in the massless quark limit when a virtual gluon gives a hard 
backward kick to one of the primary quarks.  

In our leading-order process all quark-antiquarks carry robust 
momenta except for the spectator, while in K\"{o}rner's picture 
the $q\overline{q}$ pair produced by a gluon is likely to be 
much softer. It is a dynamical question which process is more 
important to the actual baryonic decay modes through the tree 
process; the $\overline{\alpha}_s$ suppression
versus ($m/E\times$ the tail of the quark distribution in
the baryon). The purpose of our paper is to derive and test
the perturbative helicity rule for ${\cal B}\overline{{\cal B}}$
as an extension of the helicity rule in $B\to VV$. Therefore we
assume the perturbative picture and proceed to study the 
$m/E$ expansion here. 

Let us move to our argument. In the simple perturbative 
picture, three quarks fly in one direction and turn into valence 
quarks of a baryon while three antiquarks fly to the opposite 
direction and turn into valence antiquarks of an antibaryon. 
We derive the helicity selection rule on the same assumptions 
as in $B\to VV$. The chiral content of quarks and antiquarks 
is any one of the following three possibilities;
\begin{equation}
    \left\{ \begin{array}{l}
 (\overline{q_L}q_s)(\overline{q_L}q_L)(\overline{q_L}q_L)\cdots(A),\\
 (\overline{q_L}q_s)(\overline{q_L}q_L)(\overline{q_R}q_R)\cdots(B),\\
 (\overline{q_L}q_s)(\overline{q_R}q_R)(\overline{q_R}q_R)\cdots(C),
                \end{array} \right.       \label{int2}
\end{equation}
where colors are saturated separately among $qqq$ and among 
$\overline{q}\overline{q}\overline{q}$. In the standard model, 
the $(\overline{q_R}q_R)$ pair can come only from the gluon
interaction. Therefore, the final quark state (A) and (B) 
are produced by either the 
tree or the penguin interaction, but the state (C) can be 
realized only by the penguin interaction. In these final states 
the antibaryon helicity can take the value of $h=+\frac{3}{2}$ 
[$\overline{q_L}\overline{q_L}\overline{q_L}$ of (A)], 
$h=+\frac{1}{2}$ [$\overline{q_L}\overline{q_R}\overline{q_L}$ 
of (B)] or $-\frac{1}{2}$ 
[$\overline{q_L}\overline{q_R}\overline{q_R}$ of (C)]. The 
helicity of the baryon ($q_sqq$) must match the antibaryon
helicity to satisfy the overall $J_z=0$ condition. The matching 
is possible only in the case of $h=+\frac{1}{2}$ from (B) since
$h=-\frac{3}{2}$ or $-\frac{1}{2}$ for $q_sq_Lq_L$ of (A), 
$h= -\frac{1}{2}$ or $+\frac{1}{2}$ for $q_sq_Rq_L$ of (B), 
and $h=\frac{3}{2}$ or $\frac{1}{2}$ for $q_sq_Rq_R$ of (C). 
This helicity matching is shown in Table I and depicted in Fig.2.
It is easy to understand why neither the case (A) nor (C) can
satisfy $J_z=0$: When two pairs of quark-antiquark
$\overline{q_L}\overline{q_L}q_Lq_L$
(or $\overline{q_R}\overline{q_R}q_Rq_R$) fly back to back,
they are in the state of $J_z=-2$ (or +2) along the baryon
momentum. Then the remaining pair $\overline{q_L}q_s$ has
no way to turn total $J_z$ to zero.

\begin{table} [t]
\caption{Leading helicity states of the baryon and the antibaryon
which are realized by three classes of quark contents (A, B, C). 
Only $+\frac{1}{2}$ (in boldface) of the column B is compatible 
with $J_z=0$.} 

\begin{tabular}{cccc}
Hadrons   & A & B & C \\ \hline
Antibaryon & $+\frac{3}{2} (\overline{q_L}\overline{q_L}\overline{q_L})$
      & $+{\bf\frac{1}{2}} (\overline{q_L}\overline{q_R}\overline{q_L})$
           & $-\frac{1}{2} (\overline{q_L}\overline{q_R}\overline{q_R})$\\
Baryon     & $-\frac{1}{2},-\frac{3}{2} (q_sq_Lq_L)$
      & $+{\bf\frac{1}{2}},-\frac{1}{2} (q_sq_Rq_L)$
           & $+\frac{3}{2},+\frac{1}{2} (q_sq_Rq_R)$
\end{tabular}
\label{table:1}
\end{table}

\noindent
\begin{figure} [h]
\epsfig{file=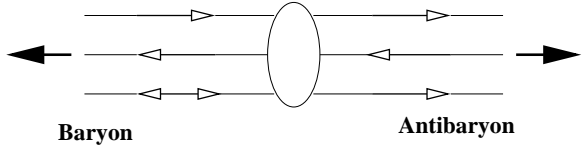,width=0.47\textwidth}
\caption{The dominant helicities of quarks and antiquarks 
($q_Lq_Rq_s +\overline{q_L}\overline{q_R}\overline{q_L}$) 
in $B\to{\cal B}\overline{{\cal B}}$ for the class (B). 
The open arrows indicate the helicity directions.
\label{fig:2}}
\end{figure}

  When the mass and transverse momentum corrections are included, 
the state of $h=-\frac{1}{2}$ is allowed for (B) with the small 
$h=-\frac{1}{2}$ component of $\overline{q_L}$ and for (C) with 
that of $q_R$. The state of $h=+\frac{3}{2}$ requires two small 
components and the state of $h=-\frac{3}{2}$ needs three small 
components. In terms of the helicity amplitudes $H_{hh}$ for 
$B\to {\cal B}({\bf p}h)\overline{{\cal B}}(-{\bf p}h)$, 
therefore, we expect most generally the hierarchy of
\begin{equation}
 \biggl|\frac{H_{-\frac{1}{2}-\frac{1}{2}}}{
       H_{+\frac{1}{2}+\frac{1}{2}}}\biggr|\approx
 \biggl|\frac{H_{+\frac{3}{2}+\frac{3}{2}}}{
       H_{-\frac{1}{2}-\frac{1}{2}}}\biggr|\approx
 \biggl|\frac{H_{-\frac{3}{2}-\frac{3}{2}}}{
       H_{+\frac{3}{2}+\frac{3}{2}}}\biggr|\approx
  \biggl\langle\frac{m_T}{E_{{\bf k}}+|{\bf k}|}\biggr\rangle ,
               \label{hierarchy2}
\end{equation}
where $m_T/(E_{{\bf k}}+|{\bf k}|)\ll 1$. For the $\overline{B}$
($b\overline{q}_s$) decay, the dominant helicity amplitude is
$H_{-\frac{1}{2}-\frac{1}{2}}$ and the hierarchy similar to 
Eq. (\ref{hierarchy2}) holds with interchange of 
$H_{hh}\leftrightarrow H_{-h-h}$. We have tabulated 
in Table II this helicity suppression for the amplitudes 
in each class ($A\sim C$) of Eq. (\ref{int2}).

 The amplitude 
of our primary interest is $H_{\pm\frac{1}{2}\pm\frac{1}{2}}$ 
since we can deparate $H_{+\frac{3}{2}+\frac{3}{2}}$
from  $H_{-\frac{3}{2}-\frac{3}{2}}$ only 
in $B\to\overline{\Omega}{\cal B}^*$. (See Section IV.) 
Since the $h=-\frac{1}{2}$ state can be realized by the small 
wrong helicity of either of two $\overline{q_L}$'s for (B), 
we may include this multiplicity factor in the ratio 
$H_{-\frac{1}{2}-\frac{1}{2}}/H_{+\frac{1}{2}+\frac{1}{2}}$;
\begin{equation}
\biggl|\frac{H_{-\frac{1}{2}-\frac{1}{2}}}{
    H_{+\frac{1}{2}+\frac{1}{2}}}\biggr|\approx  
   2 \biggl\langle\frac{m_T}{E_{{\bf k}}+|{\bf k}|}\biggr\rangle.
                     \label{hierarchy3}
\end{equation}
The approximation of $m_T\ll E_{{\bf p}}$ is a little less 
accurate for baryons than for vector-mesons since there are 
three valences instead of two and therefore the valences are
slightly less energetic.

\begin{table} [t]
\caption{The helicity factors for the amplitudes $H_{hh}$ for 
a given decay operator in the three classes $A\sim C$ of 
the final quark states. 
Here $s \equiv \langle m_T/(E_{{\bf k}}+|{\bf k}|)\rangle$ 
is the helicity suppression factor. 
The $H_{+\frac{1}{2}+\frac{1}{2}}$ amplitude of $B$ is the 
dominant helicity amplitude.  The relative
normalization between different classes depends on dynamics
as well as on the Wilson coefficients and the CKM-factors.}
\begin{tabular}{ccccc}
  $J^P$ of ${\cal B}\overline{{\cal B}}$ 
  & $h$ of $H_{hh}$ & $A$ & $B$ & $C$ \\ \hline
   $\frac{1}{2}^+\frac{1}{2}^+$ &   $+\frac{1}{2}$ &
   $4s^2$ & $2$ & $4s$  \\
                                &   $-\frac{1}{2}$ &
   $2s^2$ & $-4s$ & $4s$ \\  \hline
   $\frac{1}{2}^+\frac{3}{2}^+,\frac{3}{2}^+\frac{1}{2}^+$
    & $ \frac{1}{2}$ & $2\sqrt{2}s^2$ &
   $\sqrt{2}$ & $2\sqrt{2}s$ \\
    & $-\frac{1}{2}$ & $-\sqrt{2}s^2$    &
   $2\sqrt{2}s$ &
   $-2\sqrt{2}s$ \\ \hline
   $\frac{3}{2}^+\frac{3}{2}^+$
    & $\frac{3}{2}$ & $s^2$ & $3s^2$ & $-3s^2$ \\
    & $\frac{1}{2}$ & $2s^2$ & $1$ & $2s$\\
    & $-\frac{1}{2}$ & $s^2$ & $-2s$ & $2s$\\
    & $-\frac{3}{2}$ & $-s^3$ & $-3s^3$ & $3s^3$
\end{tabular}
\label{table 2}
\end{table}

The fraction of the helicity content 
\begin{equation}
  f_{+} \equiv \frac{|H_{+\frac{1}{2}+\frac{1}{2}}|^2}{
                     |H_{+\frac{1}{2}+\frac{1}{2}}|^2
                    +|H_{-\frac{1}{2}-\frac{1}{2}}|^2}
\end{equation}
for $B\to{\cal B}\overline{{\cal B}}$ of $J^P=\frac{1}{2}^+$
can be estimated with Eq. (\ref{hierarchy3}) when one of the 
tree or the penguin operators dominates. Following the 
approximation made in $B\to VV$, we substitute a third of the 
baryon/antibaryon momentum for the valence momentum ${\bf k}$ 
and set $m_T=\sqrt{\frac{2}{3}}\times \frac{1}{3}m_{\rm baryon}$. 
For the kinematics of the mode $B\to \overline{\Lambda}p$, 
for example, we find
\begin{equation}
   f_{+} \approx  0.89.  \;\;(\rm penguin) \label{baryon}
\end{equation} 
This is a ball-park figure for all ${\cal B}\overline{{\cal B}}$
modes. It should be reminded again that this is the number 
when a single decay operator dominates. If the dominant 
$H_{+\frac{1}{2}+\frac{1}{2}}$ amplitude consists of more 
than one term and large cancellation occurs among them, the 
value of $f_{+}$ can be smaller. However, if a large 
cancellation occurs within the $H_{+\frac{1}{2}+\frac{1}{2}}$ 
amplitude, its branching fraction would be abnormally small.
Considering the small branching fractions of the ${\cal B}
\overline{\cal B}$ modes in general, we will not be able 
to observe such abnormally suppressed 
${\cal B}\overline{{\cal B}}$ modes in the near future.

If the $\overline{b}$-quark should decay into $\overline{q_R}$ 
through the right-handed current in either the penguin or the
tree process, the $H_{-\frac{1}{2}-\frac{1}{2}}$ amplitude
would dominate in such a process according to the argument above. 
In the baryonic decay, therefore, the chirality of the weak 
current manifests itself directly in the dominant helicity 
amplitude. In contrast, the chirality of the current affects 
only the subdominant helicity amplitudes in the two-body 
meson decays.  
 
\section{Baryonic decay modes of interest}

When the baryon (antibaryon) decays by strong interactions, the 
angular correlation of the decay products with the baryon 
momentum cannot distinguish between helicity $h$ and $-h$ since 
the correlation takes the same form for $h=\pm 1$ by
parity conservation. This may look potentially a serious 
obstacle for carrying out the helicity test for the ${\cal B}
\overline{{\cal B}}$ modes. Fortunately, however, hyperons decay 
through parity-violating weak interactions and the parity 
violation can separate between helicity $\pm h$ and allow us to 
determine $f_{+}$ with a relatively small number of events.

 Let us take for concreteness the decay 
$B\to\overline{\Lambda}p$ again and choose 
$\overline{\Lambda}$ as the spin analyzer. 
The decay process is
\begin{eqnarray}
 B\to &\overline{\Lambda}({\bf p})&+p(-{\bf p})\nonumber \\
 &\searrow & \overline{p}({\bf q}) + \pi^+(-{\bf q}), 
             \label{decay}
\end{eqnarray}
where ${\bf q}$ is the decay momentum of $\overline{p}$ in the
rest frame of $\overline{\Lambda}$. Then the decay angular
distribution is given by
\begin{equation}
  \frac{d^2\Gamma}{d\Omega_{\bf p}d\cos\theta_{\bf q}}=
  \frac{\Gamma_0}{8\pi}
   (1 +\tilde{\alpha}_{\overline{\Lambda}}\cos\theta_{{\bf q}}),
\end{equation}
where $\theta_{{\bf q}}$ is the polar angle of ${\bf q}$ 
with respect to $\overline{\Lambda}$ momentum ${\bf p}$.
It is easy to show that the asymmetry 
$\tilde{\alpha}_{\overline{\Lambda}}$ is expressed with
the nonleptonic decay parameter $\alpha_{\overline{\Lambda}}$
and the helicity ratio $f_{+}$ in the form of
\begin{equation}
  \tilde{\alpha}_{\overline{\Lambda}}=
       (2f_{+}-1)\alpha_{\overline{\Lambda}}.
\end{equation}
Note that according to approximate CP invariance in the hyperon 
decay, $\alpha_{\overline{\Lambda}}= - \alpha_{\Lambda}$ holds 
to accuracy of $O(10^{-4})$ or better. 

     Since we determine the helicity amplitudes with a 
parity-violating decay, we should choose ${\cal B}$ or 
$\overline{{\cal B}}$ from the hyperons or the antihyperons 
which decay nonleptonically with large parity asymmetry. 
Therefore $\Lambda$, $\Xi$, $\Sigma^+(\to p\pi^0)$ and
their antiparticles are suitable for the spin analyzer.  
The baryon or the antibaryon that is not the spin analyzer
may be a baryon resonance, though reconstruction with too
many partcles will degrade accuracy of $f_+$.

The observation of the large transverse polarization 
in $B\to\phi K^*$ points to 
the penguin process $\overline{b}\to\overline{s}s\overline{s}$
as a primary suspect. When an additional $\overline{s}s$ pair 
is created by a gluon in $\overline{b}\to\overline{s}s
\overline{s}$, the final quark state can end up in 
$\overline{\Omega}\Xi$. Since the $\overline{s}s$ from the 
additional pair fly back to back in $\overline{\Omega}\Xi$, 
this must be a hard QCD process. Since this decay cascades down 
to six hadrons ($p\pi\pi\overline{p}\pi K$), however, it will 
not be one of the easiest modes to reconstruct. In comparison 
the decay $\overline{\Xi}\Lambda$ can be more easily studied. 
This decay occurs through either ``$\overline{b}
\to\overline{s}s\overline{s}$ (penguin) + $\overline{u}u 
(\overline{d}d)$'' or ``$\overline{b}\to\overline{u}u
\overline{s}$ (CKM-suppressed tree) + $\overline{s}s$''. 
Since the tree process is strongly suppressed by the 
CKM-factors just as in $\rho K^*$, it is safe to assume 
that $B\to\overline{\Xi}Y$ ($Y=\Lambda, \Sigma$) is dominated 
by the penguin process $\overline{b}\to\overline{s}s
\overline{s}$. 

 We thus expect that the mode $B\to\overline{\Xi}Y$ is the most 
suitable baryonic mode to study the issue raised by $\phi K^*$
in the penguin decay. 
When two nonstrange-quark pairs are emitted in the $\overline{b}
\to\overline{s}$ penguin process, the final baryon state is 
$\overline{Y}N$. This mode corresponds to $\rho K^*$ of $VV$. 
In contrast to $\phi K^*$ and $\rho K^*$, the $\rho\rho$ 
mode proceeds mainly through the tree decay 
$\overline{b}\to\overline{u_L}u_L\overline{d_L}
+\overline{q}q$ since the penguin process $\overline{b}\to
\overline{d_L}(u\overline{u}+\overline{d}d) +\overline{q}q$
is down by the loop-suppression in the Wilson coefficients
relative to the tree process. Therefore $\overline{N}N$ is
an ${\cal B}\overline{{\cal B}}$ counterpart of $\rho\rho$.
However, this mode is not useful for the polarization study 
since we need a hyperon as a spin analyzer. If one goes 
after the tree decay, an alternative is the mode 
$\overline{Y}Y$ which is dominated by the tree decay 
$\overline{b}\to\overline{u_L}u_L\overline{d_L}+\overline{s}s$. 
In short, the strangeness-changing modes ($\Delta S=1$) are 
dominated by the penguin decay while the strangeness-conserving 
modes are dominantly through the tree process
$\overline{b}\to\overline{u_L}u_L\overline{d_L}(s\overline{s})$.
This observation is not our original, but rather a consensus
among theorists in the recent papers on the subject\cite{th2}.
This simple approximation is in line with the limited numerical 
accuracy of our semiquantitative analysis. 

 With these remarks in mind, we have selected the promising
baryonic modes and listed in Table III. They are the modes 
which require reconstruction of no more than five stable 
particles and do not contain a neutron. We have not listed 
the modes that contain $\Sigma^0$ since reconstruction 
of $\Lambda\gamma$ is often difficult. Although the helicity
separation is impossible, we have included the $\overline{p}p$,
$\overline{p}\Delta^{++}$ and $\overline{\Delta}^{0}p$ modes
in the Table since they give us an idea of how large the 
branching fractions
of the interesting modes should be. The spin content 
of fast moving baryons is determined by the Lorentz-boosted 
valence quark spins, ignoring higher Fock states. We can relate 
the valence quark distributions of the octet and decuplet 
baryons with different ($I,Y$) by using the constituent 
quark model, {\em i.e.}, spin-flavor SU(6) symmetry. Then the 
baryon decay amplitudes are related to each other within 
each class ($A \sim C$) to the leading order of 
$\overline{\alpha}_s/\pi$ for short-distance QCD. 
In Table III the relative magnitudes of the dominant helicity 
amplitudes $H_{+\frac{1}{2}+\frac{1}{2}}$ are given within 
the penguin and the tree decay. Since long-distance QCD is 
included only in the baryon formation, they are more 
restrictive than the most general SU(6) symmetry prediction.  

\begin{table} [h]
\caption{The dominant baryonic decay amplitudes
$H_{\frac{1}{2}\frac{1}{2}}$ of experimental and
theoretical interest for the penguin processes
(net strangeness change $\Delta S = +1$) and the
tree processes ($\Delta S = 0$). $P$ and $T$ denote the
penguin and the tree, respectively.}
\begin{tabular}{rcc}
  &   Modes ($\Delta S= +1$)
   & Penguin ($\overline{b}\to\overline{s_L}q\overline{q}$)
                               \\ \hline
$B^0\to$ &    $\overline{\Sigma}^-p$ & $(\sqrt{6}/9)P$   \\
   &    $\overline{\Lambda}\Delta^0$ & $ 0 $  \\
   &    $\overline{\Sigma}^-\Delta^+$ & $-(2\sqrt{3}/9)P$  \\
   &    $\overline{\Xi}^0\Lambda $ & $(\sqrt{2}/3)P$  \\
   &    $\overline{\Xi}^+\Sigma^- $ & $ (5\sqrt{6}/9)P$\\ \hline
$B^+\to$  &  $\overline{\Lambda}p$ & $P$   \\
   &  $\overline{\Lambda}\Delta^+$ & $ 0$ \\
   &  $\overline{\Sigma}^+\Delta^0$ & $-(2\sqrt{3}/9)P$ \\
   &  $\overline{\Xi}^+\Lambda$ & $(\sqrt{2}/3)P$ \\
   &  $\overline{\Xi}^0\Sigma^+$ &$(5\sqrt{6}/9)P$\\ \hline
 &   Modes ($\Delta S = 0$)
   & Tree ($\overline{b}\to\overline{u_L}u_L\overline{d_L}$) \\
                       \hline
$B^0\to$ & $\overline{p}p$ & $ T$ \\
         & $\overline{\Sigma}^{*0}\Lambda$ & $-\sqrt{6}T$\\
         & $\overline{\Lambda}\Lambda, \overline{\Sigma}^{\pm}
           \Sigma^{\mp}$ & 0 \\ \hline
$B^+\to$ & $\overline{p}\Delta^{++}$ & $ -\sqrt{6}T$\\
         & $\overline{\Delta}^0 p$ & $-\sqrt{2}T$ \\ 
         &  $\overline{\Sigma}^{*0}\Sigma^+$ & $T$ \\
         & $\overline{\Lambda}\Sigma^+,
           \overline{\Sigma}^+\Lambda$ & 0
\end{tabular}
\label{table:3}
\end{table}

Apart from the obviously forbidden modes ($B\not\to
\overline{\Lambda}\Delta$ by isospin), several simple tree
decay modes ($\Delta S=0$) are absent in the leading order:
  
(i) In the tree process ($\Delta S=0$), 
$\overline{s}\overline{u_L}\overline{d_L}$ form the 
antihyperon. Therefore the antihyperon in the $\Delta S=0$ 
process is chargeless, that is, $\overline{\Sigma}^{\pm}$ 
cannot be produced.

(ii) The $\overline{u}\overline{d}$ in $\overline{\Lambda}$
is in the spin-zero state and the $\overline{\Lambda}$ spin is
carried by the $\overline{s}$ spin. Since 
$\overline{u_L}\overline{d_L}$ ($h=+1$) is in the 
spin-one state, $\overline{s}\overline{u_L}\overline{d_L}$ 
cannot form $\overline{\Lambda}$. Therefore the modes 
$\overline{\Lambda}\Lambda$ and $\overline{\Lambda}\Sigma^+$ 
are forbidden.

The magnitudes of the tree and penguin amplitudes $T$ and 
$P$ are left as free parameters in Table III. Without breaking 
up into helicity states, a few brave attempts were made in 
the past to compute the decay rates\cite{th1,th2}. 
While most agree in regard 
to the tree dominance for $\Delta S=0$ and the penguin
dominance for $\Delta|S|=1$, the ratio $|P/T|^2$ spreads
widely over two orders of magnitude depending on the methods 
and assumptions of calculation (the pole model, the diquark
model, the sum rule, {\em etc}). This large theoretical 
uncertainty is not surprising since the baryonic decay rates
depend on how the additional $q\overline{q}$ is created.
The less-known quark distribution in the baryon compounds the
uncertainty, not to mention the interference between the 
color-allowed and color-suppressed amplitudes. We give here 
only a simple order-of-magnitude estimate of $|P/T|$
based on the CKM factors\cite{PDG} and the dominant Wilson 
coefficients\cite{Buras}. In the standard notation,
\begin{equation}
 \biggl|\frac{P}{T}\biggr| \biggl( = 
   \biggl|\frac{P_{\Delta |S|=1}}{T_{\Delta S=0}}\biggr|\biggr)
  \approx 3\sqrt{6}\times \biggl|\frac{C_6}{C_2}\biggr|
      \biggl| \frac{V^*_{tb}V_{ts}}{V^*_{ub}V_{ud}}\biggr|
     \simeq  2,  \label{Wilson}              
\end{equation}   
where $3\sqrt{6}$ comes from our normalization of $P$ and $T$ 
in Table III.

From the viewpoint of branching fractions and simplicity of
analysis, the mode $B^0\to\overline{\Lambda}p$ appears to be 
the most promising among the penguin decays, while the modes
$\overline{\Sigma}^{*0}Y$ are interesting for study of the 
tree decays. The expected branching fractions for the 
$\overline{\Xi}\Sigma$ modes are as high as that of 
$\overline{\Lambda}p$. It should be noted that they are 
the values in the absence of long-distance QCD corrections
and are sensitive to dynamics too. Therefore they should not 
be taken as reliable predictions for the branching fractions.

 Although the modes listed as zero in Table III are all 
forbidden in the leading order of the perturbative picture,
they are allowed if long and/or intermediate distance strong 
interactions are enhanced or if the higher Fock configuration 
turns out to be important. That is, if substantial branching 
fractions are observed for them in future experiment, we may 
count them as an independent evidence against the perturbative 
argument.  We have also listed the modes not useful for the
helicity analysis, $\overline{p}p$, $\overline{p}\Delta^{++}$ 
and $\overline{\Delta}^0p$, since these easily identifiable
modes may give some idea about magnitude of the tree modes 
of our interest.

One major difference from the $VV$ modes is that if the 
penguin decay contains $\overline{b}\to \overline{s_R}q
\overline{q}$, this nonstandard interaction will manifest 
itself unambiguously in $f_{+}$. The ratio of 
$\overline{s_R}q\overline{q}$ to 
$\overline{s_L}q\overline{q}$ directly reflects on $f_{+}$
in the ${\cal B}\overline{\cal B}$ decays, while only 
a switch of $H_{++}\leftrightarrow H_{--}$ in the subdominant 
amplitudes occur in the $VV$ modes.
  
\section{Discussion}

  We have proposed to study the baryonic modes and collect
more information about the source of the breakdown of the 
helicity rule. Since there have already been several
proposals of possible sources, we comment on what impact 
the baryonic modes may possibly have on the issue.
  
   If the large transverse polarization of $\phi K^*$ arises 
from enhancement of the annihilation process\cite{Kagan}, the 
same enhancement is unlikely in the baryonic modes for the 
following reason: The 
annihilation decay amplitudes for $B\to VV$ are expressed with
the vector-meson form factors in the time-like region in the
leading order. They fall off like $1/q^2$ at large $q^2=O(m_B^2)$ 
in perturbative QCD, but the author of Ref. \cite{Kagan} 
suspects that the soft-collinear loop corrections enhance the 
amplitudes numerically and upset the power suppression of the 
perturbative power counting. While the baryon form factors 
similarly describe the annihilation processes into a baryon pair,
they fall off like $1/q^4$ in perturbative counting\cite{BF}. 
This difference in the asymptotic form factors can be traced
back to the dimensions of the meson and baryon wavefunctions. 
Barring the possibility that the soft-collinear loops overcome 
one more factor of $1/m_B^2$ in rate, the annihilation process 
is far less competitive in the baryon-antibaryon decay modes. 
If so, our estimate of $f_+\approx 0.9$ in Eq. (\ref{baryon}) 
should hold for most baryonic modes. If experiment disagrees
with it, we should look for other long-distance effects or
an exotic decay interaction as the cause of breakdown of the 
perturbative helicity rule.
 
If the color-magnetic decay operator of $\overline{q}
\sigma_{\mu\nu}qG^{\mu\nu}$ is responsible, as proposed in 
Ref.\cite{Hou}, conversion of the gluon $G^{\mu\nu}$ to 
$\overline{q}q$ must be enhanced to overcome the perturbative 
power suppression of $m_q/\sqrt{q^2}$ and the neutralization
of the color in $\overline{q}q$. Since the mechanism of this 
soft enhancement has not been demonstrated quantitatively, it 
is hard to extend the argument to the baryonic modes. 
Nonetheless, we can argue that such a nonperturbative 
enhancement is highly unlikely in the baryon-antibaryon decays: 
Since the $\overline{q}q$ pair originating from $G_{\mu\nu}$ 
flies back to back to form ${\cal B}\overline{{\cal B}}$, 
it is hard to avoid the short-distance chirality suppression 
of $m_q/\sqrt{q^2}\approx 2m_q/m_B$. We therefore suspect that  
enhancement of the color magnetic decay does not occur in 
the baryonic modes. We expect that $f_{+}$ should be around 
$0.9$ even if this mechanism should be responsible for the
large polarization of $VV$. If $f_{+}$ deviates largely 
from unity in ${\cal B}\overline{{\cal B}}$, a more likely
source would be a large mixture of the higher Fock 
configuration in the baryon composition.

A  proposal\cite{Kim} of the effective tensor four-quark 
interaction is ad hoc but similar to the enhanced 
color-gluon decay interaction in physical consequence.
Such an effective four-quark interaction would be suppressed 
by $m_q/\sqrt{q^2}$ in perturbative QCD, if it arises as 
a short-distance-corrected gauge interaction. Unless one 
goes outside the framework of electroweak gauge theory,
one cannot admit a large tensor interaction in any known way. 
If a short-distance tensor interaction of light quarks 
should be relevant (a long shot), it would generate the 
$H_{-\frac{1}{2}-\frac{1}{2}}$ amplitude in the 
${\cal B}\overline{\cal B}$ decay without long-distance
corrections.
     
If the large transverse polarization originates from the 
long-distance spin flip in the on-shell charmed hadron 
intermediate states\cite{Charming}, the observed effect would 
be net sum over many intermediate hadronic states. 
We have little reason to believe that a simple rule emerge 
for baryon helicity in this case. If the hadron-quark duality 
holds between the charmed-hadron-pairs and $\overline{c_L}c_L$, 
we may be able to make a crude estimate of $f_{+}$ with the 
$\langle m_T/(E_{{\cal k}}+|{\bf k}|)\rangle$ factor of the 
$c$ and $\overline{c}$-quark. In this case the values of 
$f_{+}$ for different baryonic decays would be roughly equal 
to the value of $f_L (\simeq 0.6)$ for $B\to J/\psi K^*$ 
\cite{PDG}.\footnote{
In this scenario one has to make sure that the contribution 
of the on-shell charmed-hardon intermediate states of 
the Cabbibo-suppressed tree-process does not disturb 
$f_L\simeq 1$ for the $VV$ modes such as $\rho\rho$.}
If the quark-hadron duality is not applicable, our guess is
that the values of $f_{+}$ would be statistically random 
from one baryon mode to another over a wide range centered 
around $0.5$.       

  To conclude, measurement of baryon helicity in any single $B$ 
decay mode will not decide on the source of the large transverse 
helicity observed in $B\to\phi K^*$. Nonetheless, the baryon 
helicity will be one useful additional piece of information 
not only to test the proposals so far made but also to search 
for a novel source yet unknown to us.  
  
\acknowledgments

This work was supported 
in part by the Director, Office of Science, Office of
High Energy and Nuclear Physics, Division of High Energy Physics,
of the U.S. Department Energy under Contract
DE-AC03-76SF00098, and in part by the National Science Foundation 
under grant PHY-0098840.

\end{document}